\documentclass[sigplan,10pt]{acmart}

\startPage{1}

\DeclareMathDelimiter{(}{\mathopen} {operators}{"28}{largesymbols}{"00}
\DeclareMathDelimiter{)}{\mathclose}{operators}{"29}{largesymbols}{"01}

\setcopyright{none}

\usepackage[font=small,labelfont=bf]{caption}
\usepackage[normalem]{ulem}
\usepackage{listings}
\usepackage{xparse}
\usepackage{lmodern}
\usepackage{braket}
\usepackage{qcircuit}
\usepackage{stmaryrd}
\usepackage{mathtools}
\usepackage{bussproofs}
\usepackage[capitalise]{cleveref}
\usepackage{comment}
\usepackage{amsmath}
\usepackage{graphicx}
\usepackage{amsfonts}  
\usepackage{stmaryrd}
\usepackage{arydshln}
\usepackage{appendix}
\usepackage[framemethod=tikz]{mdframed}
\usepackage{parcolumns}
\usepackage[scaled]{beramono}
\usepackage[T1]{fontenc}
\usepackage{sidecap}
\usepackage{letltxmacro}
\usepackage[bottom]{footmisc}
\usepackage{natbib}
\usepackage{mathpartir}
\usepackage{wrapfig}

\definecolor{dark-gray}{gray}{0.2}
\definecolor{mygreen}{HTML}{467E7E}
\definecolor{mygray}{rgb}{0.5,0.5,0.5}
\definecolor{mymauve}{HTML}{AC134D}
\definecolor{myblue}{HTML}{144B7D}
\definecolor{myorange}{HTML}{B25A00}

\newcommand{\RG}[1]{\textcolor{red} {RG: #1}}
\newcommand{\AJ}[1]{\textcolor{green} {AJ: #1}}

\LetLtxMacro\oldttfamily\ttfamily
\DeclareRobustCommand{\ttfamily}{\oldttfamily\csname ttsize\endcsname}
\newcommand{\setttsize}[1]{\def\ttsize{#1}}%

\newcommand{\qceq}{\push{\rule{.3em}{0em}\equiv\rule{.3em}{0em}}} 

\newcommand{\code}[1]{{\texttt{\textcolor{mymauve!70}{\bfseries  #1}}}}

\NewDocumentCommand{\codeword}{v}{%
\texttt{\textbf{\textcolor{blue}{#1}}}%
}

\DeclareCaptionLabelFormat{figure}{Fig.\nobreakspace\thefigure}
\DeclareCaptionLabelFormat{table}{Tab.\nobreakspace\thetable}
\lstdefinestyle{qubit}{
  language=Python,
  backgroundcolor=\color{white},   
  basicstyle=\linespread{0.9}\ttfamily\footnotesize,        
  breakatwhitespace=false,         
  breaklines=true,                 
  commentstyle=\color{mygreen},    
  deletekeywords={...},            
  escapeinside={<@}{@>},          
  extendedchars=true,              
  keepspaces=true,                 
  keywordstyle=\bfseries\color{myblue!70},       
  language=Octave,                 
  otherkeywords={uint, RELATIONAL_SEMANTICS, float, void, state,CLIGHTX_SEMANTICS,fresh\_int,module, class,...,import,from,def, Map, DAGCircuit ,None, assertion , terra\_call},     
  deletekeywords={get,angle, gamma, invariant},
  emph = { certiq_prove, match, end, with, let,ret, do, forall, exists, not, N,const,  list, Fixpoint, Function, Definition,Goal, Lemma, CNOT,H,Some, Rz, ForAll, And, Measure,Rx,Meas, equiv, Bloch_rep, CouplingMap, Layout, __init__, True, False, Implies, prove, BasicSwap, count_call},
  emphstyle=\bfseries\color{myorange!70},%
  showspaces=false,                
  showstringspaces=false,          
  showtabs=false,                  
  stringstyle=\color{mymauve},     
  tabsize=1,	    
 morecomment=[f][\bfseries\color{mymauve!70}][0]{@},
 morecomment=[f][\itshape\color{mygreen}][0]{//},
}

\begin{document}
\setttsize{\small}
\title{CertiQ: Mostly-automated Verification of a Realistic Quantum Compiler}

\makeatletter
\renewcommand\@date{{%
  \vspace{-\baselineskip}%
  \large\centering
  \begin{tabular}{@{}c@{}}
    First Author\textsuperscript{1} \\
    \normalsize first.author@email.com
  \end{tabular}%
  \quad 
  \begin{tabular}{@{}c@{}}
    Second Author\textsuperscript{2} \\
    \normalsize second.author@email.com
  \end{tabular}

  \bigskip

  \textsuperscript{1}Some Department, Some University\par
  \textsuperscript{2}Some Department, Some University

  \bigskip

  \today
}}
\makeatother

\author{Yunong Shi}
\affiliation{
  \institution{The University of Chicago}
}
\email{yunong@uchicago.edu}
\author{Runzhou Tao}
\affiliation{
  \institution{Columbia University}
}
\email{runzhou.tao@columbia.edu}
\author{Xupeng Li}
\affiliation{
  \institution{Columbia University}
}
\email{xupeng.li@columbia.edu}

\author{Ali Javadi-Abhari}
\affiliation{
  \institution{IBM T.J. Watson Research center}
}

\email{Ali.Javadi@ibm.com}

\author{Andrew W. Cross}
\affiliation{
  \institution{IBM T.J. Watson Research center}
}

\email{awcross@us.ibm.com}
\author{Frederic T. Chong}
\affiliation{
  \institution{The University of Chicago}
}
\email{chong@cs.uchicago.edu}
\author{Ronghui Gu}
\affiliation{
  \institution{Columbia University}
}
\email{ronghui.gu@columbia.edu}
\date{}

\thispagestyle{empty}

\begin{abstract}
    We present CertiQ, a verification framework for writing and verifying compiler passes of Qiskit, the most widely-used quantum compiler. To our knowledge, CertiQ is the first effort enabling the verification of real-world  quantum compiler passes in a mostly-automated manner. 
    Compiler passes written in the CertiQ interface with annotations
    can be used to generate verification conditions, as well as the executable code
    that can be integrated into Qiskit.
    CertiQ introduces the quantum circuit calculus to enable the efficient checking
    of equivalence of quantum circuits by encoding such a checking procedure into an SMT problem.
    CertiQ also provides a verified library of widely-used data structures, transformation functions
    for circuits, and conversion functions for different quantum data representations.
    This verified library not only enables modular verification but also sheds light on future quantum compiler design. 
   We have re-implemented and verified 26 (out of 30) Qiskit compiler passes in CertiQ,  during which three bugs are detected in the Qiskit implementation. Our verified compiler pass implementations passed all of Qiskit's regression tests without showing noticeable performance loss. 
   
 \end{abstract}
\maketitle
\section{Introduction}

 A quantum compiler is an essential component in the quantum software stack, bridging quantum hardware and useful applications. In the near-term, a quantum compiler must perform heavy optimizations on quantum programs to fit programs onto quantum devices of limited qubit lifetime and connectivity. This makes quantum compiler code error-prone, as writing correct quantum compilation transformations can be complicated.
 Bugs in quantum compilers will corrupt the execution of programs and can lead to misleading results in scientific research performed on the quantum computers (QCs). In the open-source standard, Qiskit, compiler's issue page \cite{terra_issue}, there have been numerous bugs 
  reported to date. Undetected bugs can affect the millions of simulations and real quantum machine runs executed by more than 100k users on the Qiskit platform. Thus, eliminating bugs in quantum compilers becomes a crucial problem for the success of near-term quantum computation.

Unfortunately, testing cannot provide a complete solution to the problem of quantum compiler debugging. Testing on real devices is impractical because of the noise in the hardware and the cost of state tomography. 
Testing using classical simulation is also not scalable since it requires exponential memory/time.

In theory, formal verification provides a  solution to this problem
by proving that the compiler implementation preserves the program semantics 
for source programs of any size. 
For example, in the classical case, the CompCert C compiler~\cite{Compcert} is formally verified by  constructing machine-checkable proofs in the Coq proof assistant~\cite{Coq12}. 
There have been a few efforts to verify the quantum compilation process by formal methods~\cite{Hietala2019, Amy2017, singhal2020a}. These verifiers, however, are not yet practical in real-world scenarios for three key reasons. First, these works have only shown the ability to verify  simple compiler optimizations using limited quantum data representations. However, compiler optimizations in a realistic compiler require more than these limited representations of quantum data. Second, these verifiers are written in proof assistants like Coq and F*. Thus, expertise in proof writing, familiarity with uncommon languages, and many man-hours of coding are required to conduct verification. Third, these approaches are hard to automate, making it infeasible to verify rapidly developing quantum compilers like Qiskit~\cite{terra_issue} that are heavily dependent on third-party code.



We introduce CertiQ, a verification framework 
for writing and verifying IBM Qiskit compiler optimizations. To our knowledge, CertiQ is the first effort enabling  the verification of real-world quantum compiler transformations in a mostly-automated manner. 
By ``mostly-automated,'' we mean programmers need to write few to no specifications and annotations to support verification. Though targeted at quantum verification, the high-level workflow of CertiQ is similar to classical verification frameworks such as Alive \cite{alive} and Yggdrasil~\cite{yggdrasil}:
code contributors write compiler optimizations
and few specifications in CertiQ, and 
the correctness of the code can  be
automatically verified through the CertiQ verifier based on satisfiability modulo theory (SMT) solvers.
Executable and highly-optimized Qiskit compiler passes
can then be generated directly from the 
compiler passes written in CertiQ. This makes CertiQ readily available for the 100k users of the IBM cloud service who perform over tens of millions of simulations and experiments every year.

The design philosophy underpinning CertiQ is motivated by three practical challenges that arise when
automating the verification of 
a real-world quantum compiler. The first challenge 
is that efficient quantum circuit equivalence testing is still beyond reach.
Checking quantum circuit properties
usually 
requires exponential computation time and memory,
especially when using the denotational semantics
(i.e., the matrix representation) \cite{NielsenChuang}.
The key insight behind our solution is that,
although checking general properties of quantum circuits is hard,
checking equivalence---the specific property required to verify quantum compilers---can 
be formulated as an SMT problem for an automatic checking process with
a state-of-the-art SMT solver (Z3~\cite{z3}).
Thus, CertiQ is not characterized by large amounts of computational overhead as the methodology does not depend on the full denotational semantics for verification. Instead, a set of extracted rules about provably correct circuit rewriting is used to verify quantum program transformations.
Based on this idea, CertiQ introduces a quantum circuit calculus that enables  symbolic representations and executions of quantum circuits
and offers a set of rewriting rules to safely reduce gate counts inserted by compiler passes.
CertiQ also provides a verified library containing high-level transformation functions over quantum circuits
that are proven to preserve semantics. 
Verification conditions of compiler passes written in CertiQ can be encoded into SMT formulas using
the symbolic execution results of the code as well as with theories (or prerequisites) extracted from  the
rewriting rules and  specifications of library functions.



As for the second challenge, quantum compilers greatly benefit from the ability to manipulate data in multiple representations to better facilitate optimizations,
but conversions between different representations are complicated, error prone, and hard to reason about.
CertiQ adapts the {\it forward simulation} technique~\cite{lynch1995forward}
to define the correctness of such conversions
and builds a set of verified conversion functions in the CertiQ library.
Using this procedure, we identified a critical bug in the Qiskit implementation that converts quantum data representations, which has been confirmed by the Qiskit team.


Finally, it is impractical to write and maintain two versions of the compiler code---the verified version and the executable version.
CertiQ solves this problem by designing a uniform Python-like interface to write, specify, verify, and maintain compiler passes in CertiQ,
while relying on the CertiQ synthesizer to produce two versions of the code: the symbolic representations of the compiler pass
 for the verification purpose and the executable Python code that can be integrated into the Qiskit implementation. 
This solution is based on an observation that quantum compilers are highly domain specific, where many modules and 
templates can be abstracted to allow reusability. 
For example, CertiQ abstracts three different loop invariant templates that can cover most use cases in quantum compiler passes.
We have used CertiQ to write and verify 26 (out of 30) compiler passes in  Qiskit  (version 0.192),
showing the capability of our interface and synthesizer.

This paper makes the following contributions: 
\begin{itemize}

     \item We introduce the quantum circuit calculus
        to encode the equivalence check of quantum circuits into an SMT problem
        and invoke SMT solvers to mostly automate the check.
    
    \item We have implemented and verified the CertiQ library as  the infrastructure for building new compiler passes, including basic classes for qubits and quantum gates, a set of high-level transformation functions over quantum circuits that are proven to preserve semantics, and a set of functions to safely convert quantum data representations.
    
    \item We present the CertiQ synthesizer that provides a uniform interface for writing and maintaining
    both the verified and the executable version of compiler passes.

    \item We have implemented and verified 26 out of 30 compiler passes of the Qiskit compiler using CertiQ. 
    The implementations extracted from these 26 verified compiler passes are executable on real quantum hardware
    and  have passed all of Qiskit's regression tests. There is no noticeable performance lost in the code extracted from CeritQ as compared to the original Qiskit implementation.
    
    \item We have identified three bugs in the Qiskit implementation using our verification procedure using CertiQ, two of which are specific to quantum software. 
    
\end{itemize}


\section{Background}
\label{sec:background}

In this section, we introduce the necessary background on the verification of quantum computing and quantum compilation.

\subsection{Quantum Data and Representations}
\label{sec:quantumdata}
The ability to convert quantum data between different representations provides great flexibility for quantum compilers to perform optimizations. We introduce several important representations of basic quantum data (qubits, quantum gates, quantum circuits) in quantum compilers.

\paragraph{State vector representation.}

\begin{figure}[t]
        \small
        \vspace{3pt}
        \begin{tabular}{ccc}
        \centering
           
                \begin{minipage}{0.10 \textwidth}
                \centering 
            \Qcircuit @C=0.3em @R=0.2em{
                 &\gate{X}& \qw \\
            }
            \end{minipage}
           &
            
            \begin{minipage}{0.10 \textwidth}
                \centering
            \Qcircuit @C=0.3em @R=0.2em{
                 &\gate{H}& \qw \\
            }
            \end{minipage}&

\begin{minipage}{0.10 \textwidth}
    \centering
    \Qcircuit @C=0.3em @R=0.2em{
     &\ctrl{3}  &\qw  \\
&&  \\
\\
         &\targ     &\qw  \\
}
\end{minipage}\vspace{10pt} \\

 $ \llbracket X \rrbracket = \begin{bmatrix} 0 & 1 \\ 1 & 0 \end{bmatrix}$ & $ \llbracket H \rrbracket =\frac{1}{\sqrt{2}}\begin{bmatrix} 1 & 1 \\ 1 & -1 \end{bmatrix}$ & $ \llbracket CX \rrbracket= \begin{bmatrix} 1 & 0 & 0 & 0 \\ 0 & 1 & 0 & 0 \\ 0 & 0 & 0 & 1 \\ 0 & 0 & 1 & 0 \end{bmatrix}$ 
        \end{tabular}
        \caption{ Circuit diagram symbols (top) and denotational semantics (bottom) for several 1-qubit and 2-qubit gates.}
        
        \label{fig:gatematrix}
\end{figure}
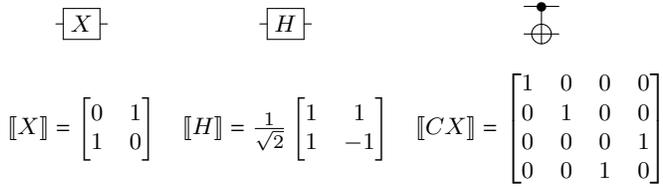

Mathematically, qubit states are $2\times 1$ normalized complex  vectors and quantum gates are unitary matrices that perform linear operations on these state vectors.  Canonically, the state vector (matrix) representation defines the {\it denotational semantics} of qubits (quantum gates) in a quantum language, providing the basis for verification. For example, for logical $\ket{0}$ and $\ket{1}$ states, $\llbracket\ket{0}\rrbracket=[1,0]^T$ and $\llbracket\ket{1}\rrbracket=[0,1]^T$, and  ~\cref{fig:gatematrix} (bottom) shows the matrix representations for the commonly used $X$ $(NOT)$, Hadamard, and Controlled-X gate. The state vector of multiple qubits is composed by performing the tensor product on individual qubit state vectors, thus scales exponentially with the number of qubits. For this reason, the state vector representation is too costly to be used in a quantum compiler beyond several qubits, and direct verification of quantum computing with denotational semantics is intractable in general.

\paragraph{Bloch sphere representation.}
\setlength{\columnsep}{8.5pt}%
\setlength{\intextsep}{0pt}%

For optimizing a sequence of single qubit quantum gates, it is often useful to map qubit state vectors to points on a unit sphere,
\begin{wrapfigure}{r}{0.13\textwidth}
    \vspace{-10pt}
  \begin{center}
    \includegraphics[width=0.13\textwidth]{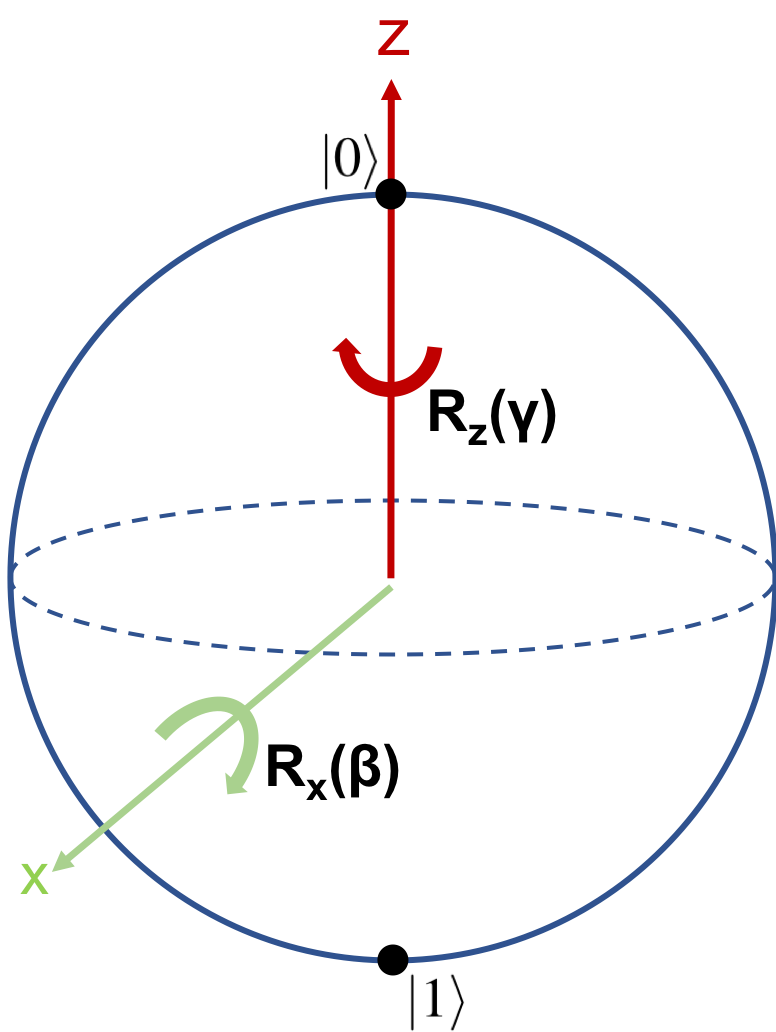}
  \end{center}
\end{wrapfigure}
 which is known as the Bloch sphere~\cite{bloch}.
 In this representation, qubit states are parameterized by their spherical coordinates $(\alpha, \beta)$ and quantum gates become fixed-axis rotations. 
For example, logical $\ket{0}$ is mapped to the north pole and logical $\ket{1}$ is mapped to the south pole and the $X$ gate becomes $180^{\circ}$ rotation with regard to the $x$-axis. The Bloch sphere representation facilitates the combination of single qubit rotations. Usually, rotations on the Bloch sphere in quantum compilers are implemented by converting values to unit Quaternions, which is a method widely adapted in classical computer graphics, $i.e.$, the underlying denotational semantics.

\begin{figure}[t]

\hspace*{1.0cm}  \begin{minipage}{0.27\textwidth}
\Qcircuit @C=0.41em @R=0.8em {
&   \lstick{\ket{0}} &\gate{H}  &\ctrl{1}   &\qw        &\qw\\
&\lstick{\ket{0}} &\qw       &\targ      &\ctrl{1}   &\qw\\
&   \lstick{\ket{0}} &\qw       &\qw        &\targ      &\qw\\
}
\end{minipage} \hspace*{1.2cm}  
\begin{minipage}{0.10\textwidth}
\centering
    \begin{verbatim}
//GHZ circuit
OPENQASM 2.0;
include "qelib1.inc";

qreg q[3];
h q[0];
cx q[0],q[1];
cx q[1],q[2];
    \end{verbatim}
\end{minipage}\hspace*{0.3cm}
\vspace{-10pt}
\caption{Circuit diagram (left) and OPENQASM IR (right) of a simple circuit.}
\label{fig:qasm}
\end{figure}
\paragraph{Intermediate representations of quantum circuits.} A quantum circuit is a sequence of quantum gates applied on an input quantum state. For general quantum circuits beyond a few qubits, it is impractical to use matrix representation or Bloch sphere representation in a quantum compiler. Thus, quantum circuits are often represented in an Intermediate Representation (IR) (see Fig.~\ref{fig:qasm}) and optimizations are performed over IR based on the compiler rules extracted from the denotational semantics. In different parts of the  compiler, different IRs can be used. For example, the Qiskit compiler uses the DAG (Directed Acyclic Graph) implementation of quantum circuits during compilation for flexible circuit manipulation and uses the OpenQASM IR for input and output. 

An important task for the verification of quantum compilers is to reason about if different representations are faithful to the underlying semantics of the quantum data. Additionally, transformations between representations must be error free. For example, the optimizations performed on any representations of quantum circuits must not violate the linear algebra rules of the unitary matrix representation.

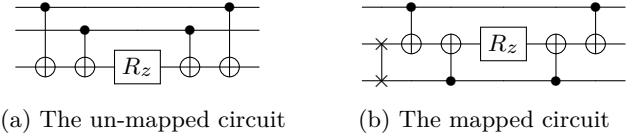
\begin{figure}[t]
    \centering \small
\hspace*{0.3cm}
    \begin{minipage}{0.30 \textwidth}
    \centering
    \Qcircuit @C=0.8em @R=0.7em {
&\ctrl{2}   &\qw        &\qw        &\qw        &\ctrl{2}   &\qw\\
&\qw        &\ctrl{1}   &\qw        &\ctrl{1}   &\qw        &\qw\\
&\targ      &\targ      &\gate{R_z}   &\targ      &\targ      &\qw
}\vspace*{0.3cm}\hspace*{-1.95cm}(a) The un-mapped circuit  
    \end{minipage}\hspace*{-0.8cm}
    \begin{minipage}{0.30 \textwidth}
    \centering
    \Qcircuit @C=0.8em @R=0.7em {
&\qw            &\ctrl{1}   &\qw        &\qw
&\qw        &\ctrl{1}   &\qw\\
&\qswap         &\targ      &\targ      &\gate{R_z}
&\targ      &\targ      &\qw\\
&\qswap\qwx[-1] &\qw        &\ctrl{-1}  &\qw
&\ctrl{-1}  &\qw        &\qw
}\vspace*{0.3cm}\hspace*{-2.1cm}(b) The mapped circuit
    \end{minipage}\\
    \vspace*{0.15cm}
    \caption{An example of circuit transformation made by lookahead swap, one of  \texttt{routing} passes. The first gate in circuit (b) is a quantum swap gate, which swaps the quantum state of the two connected qubits. Consequently, all the following gates operating on these two qubits  have to swap the two operands.}
    \label{fig:translation}
\end{figure}
\subsection{Compilation of Quantum Circuits}
\label{sec:qcompilation}
Quantum compilation is the process of translating 
high-level quantum circuit descriptions 
to optimized low-level circuits that are executable on  hardware. Most quantum compilers follow a design philosophy resembling that of LLVM~\cite{Lattner}:  circuit IRs are sequentially fed into a cascade of compiler components, called passes, to be transformed and optimized.

In the Qiskit compiler (version 0.192), there are \emph{seven} types of compiler passes: layout selection, routing, basis change, optimizations, circuit analysis, synthesis, and additional assorted passes. Layout selection passes together with routing passes make sure that the decomposed quantum circuit conforms to the topological constraints of the quantum hardware. The routing procedure is usually done by strategically inserting quantum swap gates (see~\cref{fig:translation} for an example). Basis change passes help the decomposition of quantum circuits to the gate set supported by the target hardware backend. Optimization passes include various circuit rewriting based optimizations such as gate cancellation~\cite{Maslov2008}, scheduling optimization~\cite{Shi2019}, noise adaptation~\cite{Murali2019}, and crosstalk mitigation~\cite{murali2020}. Circuit analysis passes do not modify the circuits but return important information about the circuits. Synthesis passes perform large unitary matrix decomposition. Additional passes perform miscellaneous tasks such as circuit validation.

\section{The CertiQ Workflow}
\label{sec:framework}

The goal of the CertiQ framework is to allow quantum programmers without much formal verification background
to write provably correct quantum transformation passes,
which can be readily integrated into the open-source Qiskit compiler and used in real-world quantum experiments. 

\begin{figure}[t] 
    \centering
     \includegraphics[width=0.45\textwidth]{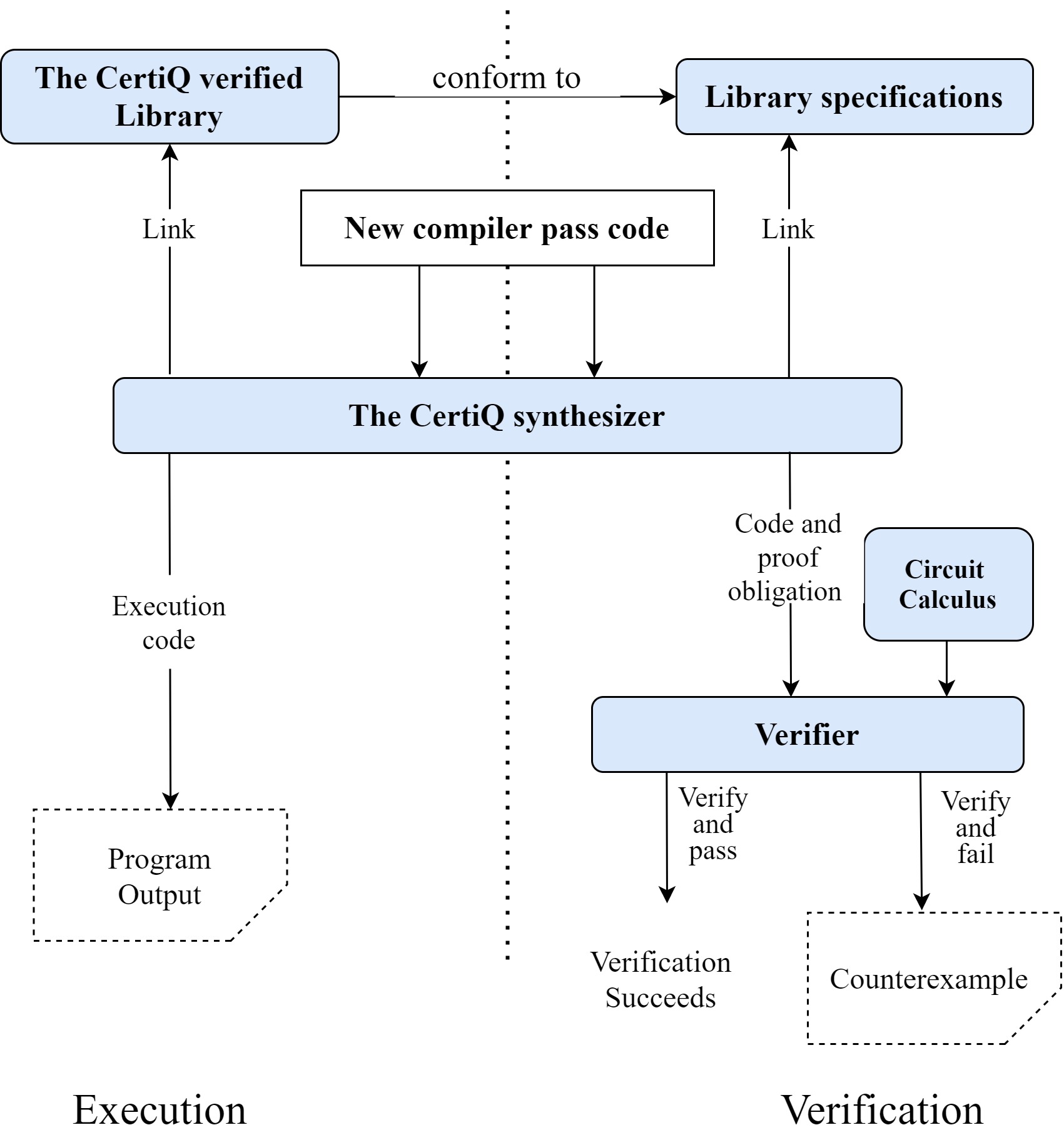}
    \caption{CertiQ workflow. Blue boxes are CertiQ components. 
    }
    \label{fig:certiq_flow}
\end{figure}

The CertiQ framework consists of five major components (see \cref{fig:certiq_flow}): 
(1) A verified library consisting of 
basic classes of quantum data,
high-level equivalent transformation functions over quantum circuits,
and conversion functions for different data representations,
serving as the basis for building compiler passes in CertiQ (see \S\ref{sec:spec});
(2) Specifications for the verified library;
(3) The quantum circuit calculus (QCC) that
introduces the symbolic 
execution of quantum circuits
with a set of pre-proven rules to reduce or rewrite
 gates inserted by compiler passes (see \S\ref{sec:equivalence});
(4) The CertiQ synthesizer, offering 
a unified interface to write and maintain
quantum compiler
passes, which can then synthesize the executable Python
 code (linked with the implementations of the library),
as well as the verification conditions (linked
with the specification of the library) 
(see \S\ref{sec:certiqlang}); 
(5) A verifier that encodes
the verification conditions into SMT formulas
and invokes Z3 to solve.

Figure~\ref{fig:certiq_flow}
shows the workflow of verifying
a compiler pass written in CertiQ.
In the rest of this section,
we will fake
a dummy compiler pass
as a running example to 
explain the verification details
involving loops, symbolic execution,
equivalence rules,
and the library.
This  dummy
compiler pass will loop over all the gates
in the circuit and, in each iteration, it first
inserts two adjacent $CX$ gates
and then applies an equivalent transformation $T$ in the library to all the following gates.

The CertiQ synthesizer first
statically analyzes the compiler pass code,
and produces symbolic representations.
The compiler pass in the running example contains a loop
and the loop iteration transfers
$C;C'$ into $C;CX;CX;T(C')$.
The symbolic representation of
the circuit after
the loop iteration is:
\begin{center}\small
$\texttt{app}(C;CX;CX;T(C'), Q)$
\end{center}
\noindent{}which applies the previous circuit $C$,
two adjacent $CX$ gates, and the transformed circuit  $T(C')$
to an input qubit register $Q$.

To prove that this pass preserves
the semantics,
we only need to show that circuits before
and after each loop
iteration are equivalent, i.e.,
\begin{center}\small
$
\texttt{app}(C;C', Q) \equiv
\texttt{app}(C;CX;CX;T(C'), Q)
$ 
\end{center}

CertiQ then performs the symbolic execution on both sides:
{\small
\begin{align*}
&\texttt{app}(C;C', Q) \rightharpoonup \texttt{app}\left(C', \texttt{app}(C,Q)\right)\\
\\
&\texttt{app}(C;CX;CX;T(C'), Q)\\
&\rightharpoonup \texttt{app}\left(CX;CX;T(C'), \texttt{app}(C, Q)\right)\\
&\cdots\\
&\rightharpoonup \texttt{app}\left(T(C'), \texttt{app}(CX, \texttt{app}(CX, \texttt{app}(C, Q)))\right)
\end{align*}
}%

\noindent{}Thus, the proof goal becomes:
{\small
\begin{align*}
{\bf G}: 
\texttt{app}\left(C', \texttt{app}(C,Q)\right)
\equiv \texttt{app}\left(T(C'), \texttt{app}(CX, \texttt{app}(CX, \texttt{app}(C, Q)))\right)
\end{align*}
}%

The quantum circuit calculus defines
a rewriting rule to cancel two adjacent $CX$ gates,
which is introduced as the following precondition to the proof goal:
\begin{center}\small
${\bf P_1}: \
\forall Q_1,\ \texttt{app}\left(CX, \texttt{app}(CX, Q_1)\right) \equiv Q_1
$ 
\end{center}

By linking with the specification of $T$ in the CertiQ library, a new precondition
about $T$ can be also introduced as:
\begin{center}\small
${\bf P_2}: \
\forall C_2\ Q_2,\
\texttt{app}\left(T(C_2), Q_2\right) \equiv
\texttt{app}\left(C_2, Q_2\right)
$ 
\end{center}

The CertiQ verifier
then encodes the proof goals and 
preconditions into SMT formulas
and invokes Z3 to check if the following
formula is satisfiable:
\begin{center}\small
${\bf P_1} \wedge {\bf P_2} \wedge \neg {\bf G}
$ 
\end{center}
\noindent{}If the above formula
is satisfiable,
the compiler pass is incorrect
and a counter-example is generated by the verifier. Otherwise when the above formula
is unsatisfiable, 
we successfully verify that the compiler pass written in CertiQ
correctly preserve the semantics.

In the next sections, we will introduce all CertiQ components  in detail. We will start with the most important, the quantum circuit calculus, that enables the efficient 
check of the equivalence of quantum circuits.

\section{Quantum Circuit Calculus}\label{sec:equivalence}
As mentioned in \S\ref{sec:quantumdata}, 
it is intractable to check the equivalence of quantum circuits  using the matrix representation (or denotational semantics) directly,
due to the exponential computational cost. 
To address this challenge,
we introduce symbolic representations and executions
for quantum circuits, and 
rules to reduce gates and perform circuit rewriting. 


Before diving into the calculus, 
we first define the syntax and semantics of quantum 
circuits
supported by CertiQ. Our syntax is similar to 
OpenQASM~\cite{cross2017open} and can be transformed to and from the DAGCircuit representation in the Qiskit  compiler.

\paragraph{Quantum circuit syntax.} The syntax of  quantum
programs (or circuits) in CertiQ can be viewed as a subset of the OpenQASM standard.
Features in OpenQASM that are not supported by  existing hardware such as classical control flow are not included in our syntax. Nevertheless, this syntax is general enough to support a wide range of gate representations, circuit transformations, and various targeting hardware. Detailed explanations of the syntax is omitted here and can be referred to OpenQASM standard~\cite{cross2017open}.
    


    \begin{figure}[h]
        \centering\small
        \begin{align*}
         \llbracket U\rrbracket_{\text{nqreg}}:=&\quad\texttt{matrix}(U_{q_1,...,q_n}) \otimes I_{\texttt{qreg} \backslash \{q_1, \ldots, q_n\}} \\
            \llbracket C_1\ ; C_2\rrbracket_{\text{nqreg}}:=&\quad\llbracket C_1\rrbracket_{\text{nqreg}} \times \llbracket C_2\rrbracket_{\text{nqreg}}
        \end{align*}
        \vspace{-10pt}
        \caption{Denotational semantics of quantum circuits and unitary operations in CertiQ. 
        \texttt{matrix} denotes the unitary matrices of the quantum operations. }
        \label{fig:denotational}
    \end{figure}
    
    \paragraph{Denotational semantics.} 
    Figure~\ref{fig:denotational} shows
    the denotational semantics of a quantum circuit $C$,
    which is defined as its corresponding unitary matrix
    and is denoted as $\llbracket C \rrbracket_{\text{nqreg}}$,
    where nqreg is the number of qubits in the quantum register used in the circuit.
    For example, the denotational semantics of an empty circuit with ${\text{nqreg}}$ qubits is the identity matrix of size ${\text{nqreg}}$. The semantics for a quantum gate is the tensor product of its matrix representation on its qubit operands and identity matrix on other unrelated qubits. 
    The semantics of the concatenation of two quantum programs is the multiplication of their matrix representations.  
    The equivalence of two quantum circuits $C_1$ and $C_2$ is then 
    defined using the equality of their denotational semantics:
    $$\forall \text{nqreg},\ \llbracket C_1 \rrbracket_{\text{nqreg}}=\llbracket C_2\rrbracket_{\text{nqreg}}$$
   


\paragraph{Symbolic representation and execution.}
A multi-qubit  quantum register $Q$ is symbolically represented as an array of symbolic qubits
$(q_1,\cdots,q_n)$.
CertiQ defines the
symbolic function
$\texttt{app}_{1q}(C, q)$
to denote the resulting qubit
of applying 
$C$ on $q$,
and defines $\texttt{app}_{2q}(C, q_1, q_2, k)$
to denote the $k$-th resulting qubit ($k\in\{1,2\}$) of applying the 2-qubit gate $C$
on $q_1$ and $q_2$. 
The result for applying the whole circuit to a symbolic quantum register is represented as \texttt{app}$(C, Q)$,
which can be executed by  applying each gate in sequence on  $Q$:
{\small
\begin{align*}
    \texttt{app}(skip, Q) &\rightharpoonup Q \\
        \texttt{app}(C_1;C_2, Q) &\rightharpoonup  \texttt{app}(C_2, \texttt{app}(C_1, Q))\\
    \texttt{app}(U(i), (q_1,  \ldots, q_n)) &\rightharpoonup (q_1, \ldots, \texttt{app}_{1q}(U, q_i), \ldots,  q_n) \\
    \texttt{app}(U(i,j), (q_1,  \ldots, q_n)) &\rightharpoonup \\ (q_1, \ldots, \texttt{app}_{2q}(&U, q_i, 1), \ldots,\texttt{app}_{2q}(U, q_j, 2), \ldots,  q_n)
\end{align*}
}%
For example, applying the GHZ  circuit
\begin{center}\small
$GHZ :=\ H(0); CX(0, 1); CX(1, 2)$
\end{center}
on the register $(q_1, q_2, q_3)$
will result in $(q_1', q_2', q_3') $ such that
{\small
\begin{align*}
q_1' &= \texttt{app}_{2q}\left(CX, \texttt{app}_{1q}(H, q_1), q_2, 1\right) \\
q_2' &= \texttt{app}_{2q}\left(CX, \texttt{app}_{2q}(CX, \texttt{app}_{1q}(H, q_1), q_2, 2), q_3, 1\right) \\
q_3' &= \texttt{app}_{2q}\left(CX, \texttt{app}_{2q}(CX, \texttt{app}_{1q}(H, q_1), q_2, 2), q_3, 2\right)
\end{align*}
}%

\begin{figure}[t]
     \centering \small
\begin{minipage}{0.30 \textwidth}
	\centering
	\Qcircuit @C=0.41em @R=0.7em {
	& \ctrl{1}    & \qw       & \ctrl{1}    & \qw      &\qw &        & &  \ctrl{2}  & \qw  \\
    & \targ       & \ctrl{1}  & \targ       & \ctrl{1} &\qw & \qceq  & &  \qw       & \qw  \\
    & \qw         & \targ     & \qw         & \targ    &\qw &        & &  \targ     & \qw 
	}
\end{minipage} \hspace*{1.0cm}
\begin{minipage}{0.30 \textwidth}
	\centering
 \Qcircuit @C=0.41em @R=0.7em {
	&     		       & \ctrl{1}    & \qw       &\ctrl{1} &\qw&      &&        & &                   &  \ctrl{2}  & \qw &\\
    & \lstick{\ket{0}} & \targ       & \ctrl{1}  &\targ  &\qw  & \qceq&&        & &  \lstick{\ket{0}} &  \qw       & \qw &  \\
    &   	 	       & \qw         & \targ     &\qw    &\qw &      &&        & & 	 		  	    &  \targ     & \qw & 	}
\end{minipage} \vspace*{0.3cm}\\\noindent\rule{8.5cm}{0.6pt}\vspace*{0.3cm}

\begin{minipage}{0.30 \textwidth}
	\centering
    \Qcircuit @C=0.81em @R=0.1em {
	& \ctrl{2}    & \ctrl{2} & \qw 
	&&  &\qw&\qw&\qw    &&  &\qw&\qw&\qw&\qw\\
	&&&&\qceq& &&&  &\qceq\\
    & \targ       & \targ    & \qw
    &&  &\gate{H} &\gate{H} &\qw    &&  &\qw&\qw&\qw&\qw\\
	}
\end{minipage} \\
\vspace*{0.3cm}\noindent\rule{8.5cm}{0.6pt}\vspace*{0.3cm}

\begin{minipage}{0.10 \textwidth}
    \centering
    \Qcircuit @C=0.3em @R=0.2em{
&\gate{Z}    &\ctrl{2}  &\qw  &&&\ctrl{2} &\gate{Z}   &\qw\\
&&& &\qceq \\
&\qw         &\targ     &\qw  &&&\targ    &\qw &\qw\\
}
\end{minipage}\hspace*{0.8cm}
\begin{minipage}{0.10 \textwidth}
    \centering
    \Qcircuit @C=0.3em @R=0.2em{
&\gate{X}    &\targ  &\qw  &&&\targ &\gate{X}   &\qw\\
&&& &\qceq \\
&\qw         &\ctrl{-2}     &\qw  &&&\ctrl{-2}    &\qw &\qw\\
}
\end{minipage}\\\vspace*{0.3cm}
\begin{minipage}{0.10 \textwidth}
    \centering
    \Qcircuit @C=0.4em @R=0.7em{
&\targ      &\qw        &\qw    &&&
&\qw        &\targ      &\qw\\
&\ctrl{-1}  &\ctrl{1}   &\qw    &&\qceq&
&\ctrl{1}   &\ctrl{-1}  &\qw\\
&\qw        &\targ      &\qw    &&&
&\targ      &\qw        &\qw
}
\end{minipage}\hspace*{0.8cm}
\begin{minipage}{0.10 \textwidth}
    \centering
    \Qcircuit @C=0.4em @R=0.7em{
&\ctrl{1}   &\qw        &\qw    &&&
&\qw        &\ctrl{1}   &\qw\\
&\targ      &\targ      &\qw    &&\qceq&
&\targ      &\targ      &\qw\\
&\qw        &\ctrl{-1}  &\qw    &&&
&\ctrl{-1}  &\qw        &\qw
}
\end{minipage}\vspace*{0.3cm}\\ \noindent\rule{8.5cm}{0.6pt}\vspace*{0.3cm}\\
\begin{minipage}{0.30 \textwidth}
	\centering
    \Qcircuit @C=0.7em @R=0.7em {
	& \qswap        &\qw & \qswap           & \qw      &     & &\qw&\qw&     &\\
	&               &    &                  &          &\qceq& &   &   &\qceq& \\
    & \qswap\qwx[-2]&\qw & \qswap\qwx[-2]   & \qw      &     & &\qw&\qw&     &\\
	}
\end{minipage}
\begin{minipage}{0.10\textwidth}
	\centering
    \includegraphics[width=0.99\textwidth]{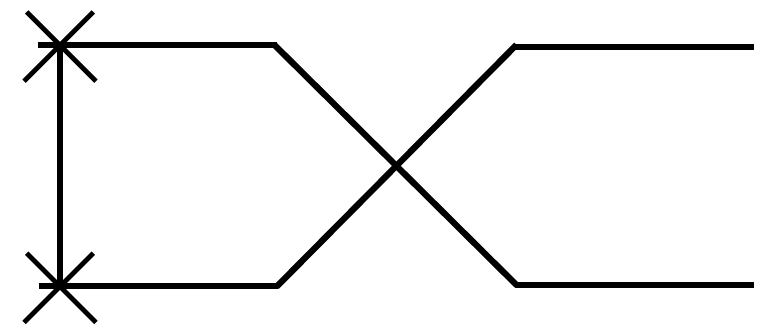}
\end{minipage}\\
    \vspace*{0.3cm}
     \caption{Examples of  rules for reducing and rewriting gates. They are bridging rules (above), cancellation rules (2nd line), commutativity rules (3rd, 4th line), and swap rules (bottom).}
     \label{fig:primitive_moves}
         \vspace*{0.3cm}
 \end{figure}

\paragraph{Rewriting rules.}
 To efficiently check the equivalence
 of symbolic representations
 of quantum circuits,
 CertiQ introduces a set of
  rules
 for reducing and rewriting gates
 that are inserted
 by compiler passes.
For example, as shown in  \cref{fig:primitive_moves},
the swap rules state that
$SWAP$ gates will swap the two qubit arguments,
which can be represented by the following 
lemmas:
{\small
\begin{align*}
\texttt{app}_{2q}(SWAP, q_1, q_2, 1) &\equiv q_2 \\
\texttt{app}_{2q}(SWAP, q_1, q_2, 2) &\equiv q_1
\end{align*}
}%
The cancellation rules applying two adjacent $CX$ gates on the same pair of qubits will not change the state:
{\small
\begin{align*}
    &\texttt{app}_{2q}(CX, \texttt{app}_{2q}(CX, q_1, q_2, 1), \texttt{app}_{2q}(CX, q_1, q_2, 2), 1) \equiv q_1 \\
    &\texttt{app}_{2q}(CX, \texttt{app}_{2q}(CX, q_1, q_2, 1), \texttt{app}_{2q}(CX, q_1, q_2, 2), 2) \equiv q_2 
\end{align*}
}%


These rewriting rules are
defined as Z3 \texttt{assertions} 
about the symbolic functions 
$\texttt{app}_{1q}$ and $\texttt{app}_{2q}$,
and are added into the list of preconditions of the proof obligations.


\paragraph{Lemmas for circuit composition.} In the verification of  compiler passes containing 
loops such as
routing passes, we have to reason about sub-circuits or the first $i$ gates of a circuit. 
CertiQ  provides 
a set of shortcut lemmas to facilitate the
compositional verification of quantum circuits
(see Appendix~\ref{sec:appendix:shortcut}).


\paragraph{Soundness proofs.}
The symbolic execution,
rewriting rules,
and the shortcut lemmas
are treated as axioms in CertiQ
and are proven using the denotational semantics
with the Coq proof assistant~\cite{Coq12}.
The soundness  of the symbolic execution
can be trivially proven by showing that
output state is always the same
 denotational semantics
with the input state.
Since all rewriting rules
only deal with a small number of gates,
their soundness  using
matrix representation can also be proven easily.
The shortcut lemmas
are proven using the compositionality and linearity of unitary matrices. 

Note that it is unnecessary to automate
these soundness proofs in the quantum circuit calculus
since they only need to be done once for all verification. 

\section{The Verified CertiQ Library}
\label{sec:spec}
    The CertiQ verified library follows the design of Qiskit to ease the learning curve for Qiskit programmers. This library implements a subset of Qiskit's functionality (without supporting pulses, schedules, etc.) that provides all infrastructure needed in writing new compiler passes, including Python classes 
    for different representations of qubits (\S\ref{sec:quantumdata}),  quantum gates, and quantum circuits, as well as a wide range of functions for circuit rewriting and representation conversion.

\paragraph{List-based  circuit implementation.} 

Unlike Qiskit's DAG implementation of quantum circuits that is heavily dependent on third party libraries, 
CertiQ implements the quantum circuit class based on generic Python lists.
With appropriate assumptions that Python lists are correctly implemented, 
we can replace Qiskit's DAG implementation (which is hard to reason about)
with CertiQ's list implementation and simplify the verification of compiler passes.

\paragraph{Verified circuit transformations.} 
Based on the basic rewriting rules
in~\cref{fig:primitive_moves}, CertiQ 
provides a library of high-level verified 
circuit transformations that can be used to 
build more complicated compiler passes. 
For example, we implement and verify the {\texttt{commutation\_cancellation}} 
transformation function in the IBM Qiskit compiler by composing the commutation rule and gate cancellation rule in~\cref{fig:primitive_moves}. 

Another example is the {\it routing} passes,
which are often implemented in a way
such that semantics preservation is broken
in the middle of the pass
and is restored later.
For example, the semantics
are not preserved right
after inserting a $SWAP$ gate
until the circuit layout is updated
for the following gates.
In CertiQ, we encapsulate the gate swapping and the layout updates in a
single transformation function, which can be verified to preserve semantics
before and after this transformation using the swap rule in~\cref{fig:primitive_moves}.
This transformation function is very useful to implement and verify
compiler passes such as those meant for {\it routing}.
The circuits in~\cref{fig:translation} provide an example. 
By invoking this transformation (shown in the following dashed box and implemented  as the \texttt{swap\_and\_update\_gate} method in the library) to the un-mapped circuit in ~\cref{fig:translation},

\begin{figure}[h]
    \centering 
    \vspace*{0.3cm}
   \begin{minipage}{0.30 \textwidth}
    \centering
    \Qcircuit @C=0.8em @R=0.7em {
&\qw&\qw                 &\qw&\qw &\ctrl{2}   &\qw        &\qw        &\qw        &\ctrl{2}   &\qw\\
&\qw&\qswap                  &\qw&\link{1}{-1}   &\qw     &\ctrl{1}   &\qw        &\ctrl{1}   &\qw        &\qw\\
&\qw&\qswap\qwx[-1]  &\qw&\link{-1}{-1} &\targ      &\targ      &\gate{R_z}   &\targ      &\targ      &\qw
\gategroup{2}{3}{3}{5}{1em}{--}
}
    \end{minipage}
    
    \vspace*{0.3cm}
    \label{fig:swap}
\end{figure}

 \noindent{}we can generate the mapped circuit in~\cref{fig:translation} without 
 exposing the intermediate inequivalence of quantum circuits to the level of symbolic executions. Thus, this verified transformation function 
 enables the automated verification of {\it routing} passes. 

\paragraph{Verified quantum data conversions.} 

Converting quantum data between different  representations enables powerful quantum circuit optimizations in the compiler. The verification
of quantum compilers must also provide a correctness guarantee for these data conversions as they are much less intuitive and more error prone as compared to the classical case. The CertiQ compiler provides quantum data converters for basic quantum data structures, including conversions between the list-based circuit implementation and
the Qiskit circuit implementation
and conversions between the qubit IR representation and the Bloch sphere representation.

To verify the correctness of such conversions, we have to first define the \emph{conversion relation}
(denoted as $\sim$)
between different data representations.
This conversion relation does not have to be 1-to-1 correspondence. For example, 
we can say that  a list implementation $C_l$  and a DAG implementation $C_d$ are convertable, i.e.,
$C_l \sim C_d$, 
if
$C_l$ and $C_d$ have the same quantum registers and $C_l$ is a topological sorting of $C_d$.

We can then use forward simulation~\cite{lynch1995forward}
to show that the relation $\sim$ is valid
and the conversion is sound.
Two representations
$D_a$ and $D_c$ of the same data $D$
can be safely converted
if and only if
the simulation diagram holds
as shown in \cref{asplos21-templates/fig/contextual_refinement}.
The simulation diagram says that,
starting from two related states
$D_a$ and $D_c$,
the resulting states $D_a'$ and $D_c'$
by applying any valid transformation $t$
must still be related.
Note that the transformation $t$
 may be implemented differently
for different representations.


\begin{figure}[h]
    \centering
    \includegraphics[width=0.19\textwidth]{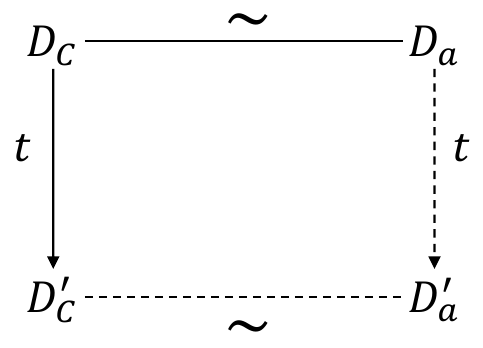}
    \caption{The simulation diagram defines the condition in which data representations
    can be converted safely.  }
    \label{fig:contextual_refinement}
    \vspace*{0.3cm}
\end{figure}

Take the conversion between the Bloch sphere representation and the qubit state vector representation as an example.
For single qubit optimizations, IRs of qubits (and 1-qubit gates) are often expanded to the state vector (matrix) representation. Further, it will be transformed to the Bloch sphere representation.
Specifically, Bloch sphere representation $(\theta, \phi)$ (see \S\ref{sec:quantumdata}) is a projection of a general qubit state vector $\ket{\psi} = e^{i\gamma}\text{cos}(\theta/2)\ket{0} + e^{i\gamma}e^{i\phi}\text{sin}(\theta/2)\ket{1}$, where the global phase $\gamma$ of a qubit state $\ket{\psi}$ is omitted. 
If we use this projection as the conversion relation, 
this relation can be written as the following Python code:
\begin{center}
\begin{tabular}{c}
\hspace*{-0.2cm}
\begin{lstlisting}[style=qubit]
def Bloch_rep(gamma, theta, phi): return (theta, phi)
\end{lstlisting}
\end{tabular}
\end{center}
\noindent{}However,  the simulation diagram in \cref{fig:contextual_refinement}
does not hold for this relation when the transformation $t$ is \texttt{tensor\_product} or any other multi-qubit operations.
The simulation diagram breaks because the untracked phase $\gamma$ will induce a 2-qubit phase gate (Controlled Z rotation) between qubits beyond the 1-qubit case. This relative phase change will induce non-trivial quantum computation that is not revealed in the Bloch representation. To address this issue, in CertiQ, we explicitly exclude conversions between the Bloch sphere representation and other representation for cases beyond 1 qubit. CertiQ used this conclusion to detect a 
severe bug in Qiskit (see \S\ref{sec:op_1q_pass}).

\section{The CertiQ Synthesizer}
\label{sec:certiqlang}
  The CertiQ synthesizer aims to provide language-level support for writing,
 specifying, and maintaining compiler passes. It consists of two components: a parser that parses the annotated user code into verification conditions and a wrapper library for user code to run in Qiskit. 

\subsection{Generating Verification Conditions}
The CertiQ interface is a Python-like language (supporting branches, loops, and function calls) 
for writing compiler passes with annotations
(see Appendix~\ref{sec:appendix:interface}).
Because the SMT solver in the CertiQ verifier does not support branch statements and loops  that are widely used in compiler passes, the  synthesizer has to  parse
the annotated code into 
 verification conditions (VCs) that are acceptable by the  verifier.
\paragraph{Branch statements.}
The CertiQ synthesizer expands all branch statements.
The verification condition 
of a branch is expanded
into two separate VCs: one for each branch.
The ``true'' branch condition will be added to the list
of preconditions for a ``true'' branch implementation while  negation of the branch condition will be added to the list
of preconditions
 for the ``false'' branch. These preconditions are then used to generate further VCs. Note that CertiQ requires that all branch conditions must be representable as SMT formulas to enable
the verification.

Although this expansion approach may lead to an exponential number of VCs, fortunately, the number of branches remains tractable in Qiskit compiler passes. For all 26 passes that we have verified, the number of branches is at most five.

\paragraph{Loop statements.}
Loops are unavoidable in compiler implementations, but SMT solvers cannot handle variable-length loops. Thus, we must provide additional information, such as  loop invariants.
Fortunately, in this domain-specific environment (a quantum compiler), this problem can be easily and practically solved as loops in quantum compilers often follow fixed patterns. CertiQ provides \emph{three} loop templates as library functions that take the loop body as an argument. We found these three loop templates can cover most of the Qiskit compiler passes. For example,  \texttt{iterate\_all\_gates(circ, func)} is
one template for loops that iterates over all the gates in \texttt{circ} and applies the same loop body \texttt{func} to each gate.

To synthesize such a loop template
from the original implementation,
CertiQ requires users to specify
the variable name in the  loop body that should be mapped 
to the \texttt{circ} argument of 
\texttt{iterate\_all\_gates(circ, func)}:
\begin{lstlisting}[style=qubit]
 for gate in qcirc.gate_list:
     #@ circ: new_qcirc
     ... # loop body generates new_qcirc with gate
\end{lstlisting}

Loop templates provided by CertiQ pre-define
loop invariants as pre- and postconditions about the loop body. 
In the below \texttt{iterate\_all\_gates} example, this loop template maintains an optimized version of a prefix of the original circuit while adding new gates one by one. The invariant should be that the circuit referred to as \texttt{new\_circ} within the loop should be equivalent to the first $i$ gates in the original circuit at $i$-th iteration of the loop. This loop template
is written as follows:
\begin{lstlisting}[style=qubit]
def iterate_all_gates(circ, func):
    # subgoal
    assertion.push()
    i = Int("i")
    n = circ.size()
    cur_circ = QCircuit()
    assertion(i >= 0)
    assertion(i + 1 < n)
    assertion(equivalent_part(cur_circ, circ, i))
    new_circ = func(cur_circ, circ[i])
    certiq_prove(equivalent_part(new_circ, circ, i+1))
    assertion.pop()
    
    ret_dag = DAG()
    assertion(
        equivalent_part(ret_circ, circ, circ.size()))
    return ret_circ
\end{lstlisting}
We can see that \texttt{iterate\_all\_gates()} defines loop invariants as pre- and postconditions
of the loop body stored as \texttt{func}. In the verification process, when CertiQ invokes this loop template, it will generate and try to prove the subgoal that when the current \texttt{new\_circ} (represented by \texttt{cur\_circ} in the specification) is equivalent to the first $i$ gates of \texttt{circ} before the $i$-th iteration, the  \texttt{new\_circ} after this iteration must be equivalent to the first $i+1$ gates of \texttt{circ}. The specification states that under the preconditions, the result of the loop is a circuit that is equivalent to \texttt{circ}.

The other two loop templates provided by CertiQ are  (1) \texttt{while\_gate\_remaining()} that is used in transformation passes such as \texttt{lookahead\_swap}, and (2) \texttt{collect\_runs()} that is used in passes such as  \texttt{optimize\_1q\_gate} (\S\ref{sec:op_1q_pass}).
Together with \texttt{iterate\_all\_gates}, these three templates help CertiQ prove all loops
in the 26 passes  of Qiskit.

\paragraph{External function calls.}
For third party libraries and complicated functions using unsupported language features or data structures, CertiQ  allows 
the programmer to label the function as an external function and specify its behavior using the ``@external'' decorator. 
The  synthesizer will
add the specification of an external function
to the precondition list
of programs invoking that external function.
External functions will not be verified by CertiQ
with respect to their specifications
and, thus, are in our trusted computing base (TCB).

\subsection{Generating Execution Code}

In CertiQ, compiler passes rely on the verified CertiQ library, thus not directly compatible with the Qiskit framework. To generate compiler passes that can be integrated into Qiskit, CertiQ provides  a Qiskit wrapper for CertiQ compiler passes that utilizes the data conversion library provided by the verified  CertiQ library. The Qiskit wrapper performs the following steps to incorporate CertiQ passes into the Qiskit compilation process: 1) it converts the input DAG circuit from the Qiskit compilation flow to the OpenQASM IR; 2) then it invokes the CertiQ pass and receives the compiled circuit from the CertiQ pass;  3)  it converts the compiled CertiQ circuit  to the corresponding DAG circuit. With these steps, the Qiskit wrapper allows CertiQ integration into Qiskit.

\section{The CertiQ Verifier}
\label{sec:verifier}
The CertiQ verifier generates proof obligations
in the form of SMT formulas and invokes
Z3 to automate the proof.

\paragraph{Generating proof obligations.}
In CertiQ, users do not need to explicitly 
provide specifications to  the verifier as in other automated verification frameworks such as Yggdrasil~\cite{yggdrasil}. 
Instead, CertiQ will load the pre-defined 
 proof obligations of all seven types of compiler passes.
For example, for all six types of compiler passes (excluding the analysis passes), 
we specify a proof obligation that the input and output circuits are equivalent through the \texttt{test()} method:

\begin{center}
\begin{tabular}{c}
\begin{lstlisting}[style=qubit]
class TransformationPass():
    @classmethod
    def test(cls):
        optimizer = cls()
        init_circ = QCircuit()
        out_circ = optimizer.run(init_circ)
        certiq_prove(equivalent(out_circ, init_circ))
        print(cls.__name__ + " verified")
\end{lstlisting}
\end{tabular}
\end{center}

\noindent{}In the above \texttt{test()} method, CertiQ first generates symbolic circuits \texttt{init\_circ} to represent the input
quantum circuits, then symbolically executes the pass implemenation
through \texttt{optimizer.run}, and finally attempts to verify the proof obligations
of circuit equivalence. When users implement a customized, device-independent transformation pass, they can use the \texttt{TransformationPass} virtual class as parent class
to guide the generation of proof
obligations:
\begin{center}
\begin{tabular}{c}
\begin{lstlisting}[style=qubit]
class MyOwnPass(TransformationPass):
    #implementation omitted
    #...
\end{lstlisting}
\end{tabular}
\end{center}

\paragraph{Verification engine.}
Our verification engine is similar to that of Alive~\cite{alive} and Yggdrasil~\cite{yggdrasil} and maintains a \emph{stack} of proof obligations. One difference is that CertiQ uses ArrayEx (Theory of Array) but Alive and Yggdrasil uses BV or QF\_BV (quantified/quantifier-free bitvector) theories.
A proof obligation is a precondition list paired with 
a claim list
meaning
that the list of claims should hold
under the list of preconditions.
At the beginning, 
the preconditions and the claims
of a function will be pushed onto the stack as a proof obligation.
The CertiQ verification engine will keep popping the top proof obligation from the stack
until it becomes empty.
The claims of a function will automatically unfold
through symbolic execution.
When executing a specification,
the engine will add the guarantees in the specification to the precondition list of the 
current proof obligation.
If the specification contains preconditions,
the verification engine will create 
a new proof obligation where the precondition list is the current precondition list,
and the claim list holds this precondition of the specification.
Such a new obligation will then be 
pushed into the stack.
Once a proof obligation is fully unfolded,
the verification engine will use Z3 to check whether the list of claims  holds 
under the list of preconditions.


\section{Case Studies}
\label{sec:cases}

After introducing all of the technical details, we present two case studies to show how the CertiQ framework can detect quantum-specific bugs in the Qiskit compiler. 

 \subsection{The \texttt{optimize\_1q\_gate} pass}
 \label{sec:op_1q_pass}
We first focus on the verification of the \texttt{optimize\_1q\_gate} pass and show that, using CertiQ, we can reveal bugs that only arise in quantum software.

We verify the re-implemented  \texttt{optimize\_1q\_gate} pass using the \texttt{merge\_1q\_gate} method. 
This pass collapses a chain of single-qubit gates into a single, more efficient gate~\cite{McKay2018}, to mitigate noise accumulation. It operates on $u_1, u_2, u_3$ gates, which are native gates in the IBM devices. 
These gates can be naturally described as linear operations on the Bloch sphere, for example, $u_1$ gates are rotations with respect to the Z axis. For clarity, we list their matrix representations in \cref{tab:gates}. 

\begin{table}[h] 
    \vspace*{0.3cm}
    \centering \small
    \begin{tabular}{l  c  r}
    $u_1(\lambda) = \begin{pmatrix}1&0\\0&e^{i\lambda}\end{pmatrix}$, &  \multicolumn{2}{c}{$u_2(\phi, \lambda) = \frac{\sqrt{2}}{2} \begin{pmatrix}1&-e^{i \lambda}\\e^{i \phi}&e^{i(\lambda+\phi)}\end{pmatrix}$} \\[0.55cm] \hline \\
    \multicolumn{3}{c}{$u_3(\theta, \phi, \lambda) = \frac{\sqrt{2}}{2} \begin{pmatrix}\text{cos}(\theta)&-e^{i \lambda}\text{sin}(\theta)\\e^{i \phi}\text{sin}(\theta)&e^{i(\lambda+\phi)}\text{cos}(\theta)\end{pmatrix}$}\\[0.55cm] 
    \end{tabular}
    \caption{Matrix representation of physical gates $u_1$, $u_2$ and $u_3$. $u_1$ is a Z rotation on the Bloch sphere.}
    \label{tab:gates}


\end{table}

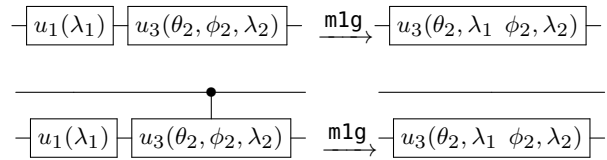
\begin{figure}[h]
    \centering \small
       \vspace*{0.4cm}
    \hspace*{-0.1cm}
    \vspace*{0.3cm}
    \hspace*{-0.22cm}
    \Qcircuit @C=.6em @R=.3em @!R {
    & \gate{u_1(\lambda_1)} & \gate{u_3(\theta_2, \phi_2, \lambda_2)} & \qw && &\xrightarrow{\texttt{\lstinline{m1g}}}&&&
    \gate{u_3(\theta_2, \lambda_1+\phi_2, \lambda_2)} & \qw
    }\vspace{.2cm}
    \Qcircuit @C=.6em @R=.3em @!R {
    &\qw &\ctrl{1}&\qw
    &&&&&
    &\qw&\qw\\
     & \gate{u_1(\lambda_1)} & \gate{u_3(\theta_2, \phi_2, \lambda_2)} & \qw && &\xrightarrow{\texttt{\lstinline{m1g}}}&&&
    \gate{u_3(\theta_2, \lambda_1+\phi_2, \lambda_2)} & \qw    } 
    \vspace*{0.55cm}
      \caption{Correct execution (top) and incorrect execution  (bottom) of \texttt{\lstinline{merge\_1q\_gate}}.}
    \label{fig:opt1qpass}
\end{figure}

The \texttt{optimize\_1q\_gate} pass has two function calls. First, it calls the \texttt{collect\_runs} method to collect groups of consecutive $u_1$, $u_2$, $u_3$ gates. Then it calls \newline \texttt{merge\_1q\_gate} to merge the gates in each group. \texttt{merge\_1q\_gate} (\cref{fig:opt1qpass}) first transforms the single qubit gates from the Bloch sphere representation to the unit quaternion representation \cite{Gille2009}, then the rotation merges are performed in that representation. 

As described in \cref{sec:spec}, conversion between these two representations allow only single-qubit operations to adhere to the simulation diagram. However, in Qiskit, all gates can be modified with a \texttt{c\_if} or \texttt{q\_if} method to condition its execution on the state of other classical or quantum bits. When the transpiler pass attempts to optimize these conditional gates, it can lead to an incorrect circuit. For this reason, in the implementation of this pass, we have to include that \texttt{gate1.q\_if == }\code{False} and \texttt{gate1.c\_if == }\code{False} unless both are controlled and the control bits are the same. 
 
  The bug described above, which relates to how quantum circuit instructions can be conditioned, have been observed in Qiskit in the past~\cite{qiskitbug1,terra_issue}. In the absence of rigorous verification like in this work, such bugs are  hard to discover. In practice, this is usually done via extensive randomized testing of input and output circuits, which does not provide any guarantee of finding faulty code.
 The results here  demonstrate that our {\it forward simulation} technique for \texttt{merge\_1q\_gate} is effective for detecting quantum-related bugs. 

\subsection{\texttt{commutation} passes}
\label{sec:cases_comm}
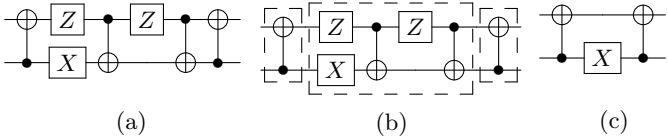
\begin{figure}[h]
    \centering \small
    \vspace*{0.4cm}
    \hspace*{-0.2cm}
    \begin{minipage}{0.30 \textwidth}
    \centering
\Qcircuit @C=.5em @R=.5em @!R{
    &\targ  &\gate{Z} &\ctrl{1} &\gate{Z} &\ctrl{1} &\targ &\qw \\
    &\ctrl{-1} &\gate{X} &\targ &\qw &\targ &\ctrl{-1} &\qw
    }\vspace*{0.4cm} \hspace*{-2cm}(a) 
    \end{minipage}\hspace*{-2.0cm}
    \begin{minipage}{0.30 \textwidth}
    \centering
        \Qcircuit @C=.5em @R=.5em @!R{
    &\targ  &\qw&\gate{Z} &\ctrl{1} &\gate{Z} &\ctrl{1} &\qw&\targ &\qw \\
    &\ctrl{-1}&\qw &\gate{X} &\targ &\qw &\targ &\qw&\ctrl{-1} &\qw
    \gategroup{1}{2}{2}{2}{.7em}{--}
    \gategroup{1}{4}{2}{7}{.7em}{--}
    \gategroup{1}{9}{2}{9}{.7em}{--}
    }\vspace*{0.2cm}\hspace*{-2cm} (b) 
    \end{minipage}\hspace*{-1.7cm}
    \begin{minipage}{0.30 \textwidth}
    \centering
    \Qcircuit @C=.5em @R=.5em @!R{
    &\targ &\qw &\targ &\qw \\
    &\ctrl{-1} &\gate{X} &\ctrl{-1} &\qw
    }\vspace*{0.45cm}\hspace*{-3.5cm} (c) 
    \end{minipage}\\\vspace*{0.1cm}
    \caption{A working example of  \texttt{commutation\_analysis} and \texttt{commutative\_cancellation}. (a) The un-optimized circuit, (b) \texttt{commutation\_analysis} forming the commutation groups, and (c) \texttt{commutative\_cancellation} cancels self-inverse gates inside groups.}
    \label{fig:comm_pass}
   
\end{figure}

\texttt{commutation\_analysis} and \texttt{commutative\_cancellation} are a pair of compiler passes that optimize Qiskit DAGCircuits using the quantum commutation rules and the cancellation rules pictured in \cref{fig:primitive_moves}. First, \newline \texttt{commutation\_analysis} transforms the quantum circuit to a representation called commutation groups \cite{Shi2019} where nearby gates that commute with each other are grouped together. Next, \newline \texttt{commutative\_cancellation} performs cancellation inside the newly formed groups. In \cref{fig:comm_pass}, we give a working example.

We find two bugs when re-implementing these two passes. First, the commutation group can be viewed as another representation of the quantum circuit. However, when we try to convert the symbolic quantum circuit representation to the commutation group representation defined in the \texttt{commutation\_analysis}, we find that the commutation group representation does not satisfy the simulation diagram defined in 
\cref{sec:spec}. We found this violation comes from the fact that the commutation relation is in general not transitive. For example, if we denote the commutation relation as $\sim$ and there is 3 quantum gates, $A, B, C$ where $A\sim B$, $B\sim C$, then $A\sim C$ is not guaranteed to be true. For this reason, gates with pairwise commutation relations cannot be grouped together. We propose two solutions to this bug. First, we can make sure the circuits that these passes operate on  have a limited gate set where $\sim$ is indeed transitive. For example, in the gate set \{CX, X, Z, H, T, $u_1, u_2,u_3$\}, $\sim$ is transitive. Second, we can use a new algorithm that does not assume transitivity.

\subsection{\texttt{routing}  passes}
\label{sec:cases_swap}

For routing passes, there are three proof obligations for \texttt{swap} passes:  
\begin{itemize}
    \item The pass must be semantics preserving.
    \item The output DAGCircuit of the pass must conform to the coupling map of the physical device.
    \item The pass must terminate.
\end{itemize}

The first proof obligation is same with all other passes. For the second proof obligation that ensures the pass correctly accomplishes its goal of transforming the circuit to match device coupling constraints, we must verify that every 2-qubit gate in the output circuit operates on two neighboring physical qubits. 
For the third proof obligation, CertiQ does not attempt to find a complete solution because it is undecidable. Instead, CertiQ aims to provide sound termination analysis for practical implementations. First, CertiQ concretizes the problem to verify the termination of passes with an input circuit of bounded depth on a given coupling map by switching to the bitvector model in the Z3 SMT solver. Termination can be proved by constructing strictly monotonic functions in a finite domain. For program states that are not in a loop or a recursive function, the program counter is a monotonic function to provide a termination guarantee. For variable-length loops and while loops, CertiQ allows users to provide a monotonic function, for example,
\begin{center}
\begin{tabular}{c}
\begin{lstlisting}[style=qubit]
gates_remaining = circ.gate_list()
while gates_remaining != 0:
# @mono: -gates_remaining.size
  ... # implementation code
\end{lstlisting} 
\end{tabular}
\end{center}
Then in the backend SMT solver in the verifier solves for the circuit input that keeps \texttt{gates\_remaining.size} unchanged and gives it as a counter example.

\begin{figure}[h]
    \centering\small
    \vspace*{0.2cm}
    \hspace*{0.3cm}\begin{minipage}{0.30 \textwidth}
    \Qcircuit @C=0.8em @R=0.3em{
  &\lstick{\scriptstyle{Q_0}}  &\ctrl{1} &\qw        &\ctrl{3}   &\qw\\
  &\lstick{\scriptstyle{Q_8}}  &\targ    &\ctrl{1}   &\qw        &\qw\\
  &\lstick{\scriptstyle{Q_7}}  &\ctrl{1} &\targ      &\qw        &\qw\\
  &\lstick{\scriptstyle{Q_{14}}} &\targ    &\qw        &\targ      &\qw
    }
    \end{minipage}\hspace*{0.5cm}
    \begin{minipage}{0.30 \textwidth}
    \includegraphics[width=1.0\textwidth]{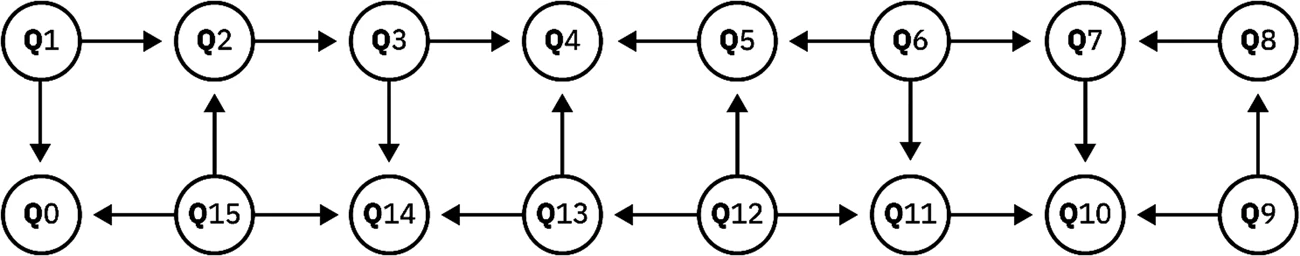}
    \end{minipage}
    \vspace*{0.15   cm}
wwwww    \caption{(left) A counter-example generated by CertiQ that shows Qiskit's \texttt{lookahead\_swap} pass does not always terminate on the IBM 16 qubit device. (right) The coupling map of the IBM 16 qubit device. Arrows indicate available CNOT directions (which does not affect the swap insertion step).}
    \label{fig:counter}
\end{figure}

We verified two routing passes: \texttt{basic\_swap}, \newline \texttt{lookahead\_swap}. We report that all two passes comply with the first two proof obligations. However,  we find a counter-example circuit on coupling map of the IBM 16 qubit device, where the \texttt{lookahead\_swap} pass does not terminate on (see \cref{fig:counter}). The \texttt{lookahead\_swap} pass greedily finds the next best 4 \texttt{swap} gates to minimize the total distance of the unmapped 2-qubit gates. However, the counter example we found shows that the 4 \texttt{swap} gates can cancel through applying the swap rules in \cref{fig:primitive_moves} and \texttt{gate\_remaining.size} will not update. This bug can be fixed by introducing randomization to break ties when needed.

\section{Evaluation}
\label{sec:evaluation}
In CertiQ, we implemented  26 out of 30 compiler passes from Qiskit.  For several passes, such as routing passes, we wrote two versions, one calling the verified CertiQ transformation library to rewrite circuits, the other directly modifying the input circuits. 
 All implementations of the 26 passes  can be verified successfully and all the generated Qiskit passes passed the regression tests provided by Qiskit. The 30 passes are listed in Qiskit v0.192 website (the number becomes 42 passes in a recent update of Qiskit v0.200 during the submission).  Among all 30 passes,  4 passes that we do not verify  are: {\it StochasticSwap}, {\it FixedPoint}, {\it DAGFixed Point}, {\it CrosstalkAdaptiveSchedule}, which either include data structures we cannot support or are meta-optimizations. For example, {\it FixedPoint} is a compiler pass that controls the execution of another compiler pass and repeatedly executes that pass until two consecutive compilation gives the same result (thus, reaches a ``fixed point'').
We evaluated CertiQ based on the implementions of the 26 verified passes.


\paragraph{Automation level.}
CertiQ is mostly automated: when writing CertiQ compiler passes, the only additional verification
effort is to write annotations for generating  loop invariants (not the loop invariants themselves),
as well as for external functions. Programmers will have to learn the CertiQ interface, which, however, is a limited burden to programmers who are already familiar with Qiskit. 

\paragraph{Verification performance.} 
For all compiler passes written with high-level transformation functions in the CertiQ library,
their verification can finish within seconds.
For passes that only apply low-level rewriting rules, 
their verification can be much slower than the ones using  using high-level transformation functions
but can still finish within one minute.

\section{Related work}
\label{sec:related}

\paragraph{Quantum programming environments with a verifier.}
Several quantum programming environments support the verification of quantum programs running on them. For example, in the QWire quantum language  \cite{Rand2018, Rand}, programmers can use the embedded verifier based on the Coq proof assistant \cite{Coq12} to create mechanized proofs for their programs. The $Q\ket{SI}$ programming environment \cite{Liu2018} allows users to reason about their programs written in the quantum {\bf while}-language with quantum Floyd-Hoare logic \cite{Ying2011}. 
In contrast to CertiQ, these environments require expertise both in quantum computing and 
mechanized verification to manually write proofs in proof assistants.
Besides, these verifiers are built for  verifying  quantum programs
rather than compiler passes to transform quantum programs.

\paragraph{Verified quantum-related compilers.} Previous studies on
compiler verification for reversible circuits \cite{Amy2017}, ancillae uncomputation \cite{Rand} and compiler optimizations \cite{Hietala2019} utilize interactive theorem provers such as F* \cite{fstar} and Coq \cite{Coq12} to conduct manual proofs,
which do not provide an extensible interface for developers
to verify future extensions with a low proof burden.
In contrast, the CertiQ verification framework allows developers to
write  new compiler passes and automate their verification.


Algorithms to perform efficient quantum circuit equivalence checking have been discussed from the view of quantum algorithms \cite{Viamontes2007}, quantum communication protocols \cite{CarlosGarcia2011}, and verification of compilation \cite{Amy2019}.  However, while powerful,  these  checking algorithms 
are too complicated for use in automated manner like what is possible with CertiQ verification.
\paragraph{Push-button verification in classical computing.} Push-button verification has been applied to building  verified compilers,  file systems, OS kernels, and system monitors~\cite{yggdrasil,hyperkernel, serval}. Our technical choice and many of the verification ideas are heavily influenced by these previous works. However, these systems cannot support variable-length loops,
calculus of quantum circuit equivalence, and
our contract-based reasoning.

\section{Conclusion}
\label{sec:conclusion}
We presented CertiQ, a verification framework that, for the first time,
enables the mostly automated verification for real-world quantum compiler passes. 
CertiQ introduces the quantum circuit calculus to efficiently check
the equivalence of quantum circuits
and provides a verified library for writing new compiler passes and modularizing their proofs.
Using CertiQ, we have successfully verified 26 (out of 30) passes of  Qiskit,
 the most widely used Quantum compiler,
 and detected several  quantum-specific bugs in the original Qiskit implementation.
The approach we establish with CertiQ paves the way for end-to-end verification of a complete quantum software toolchain, an important step towards practical near-term quantum computing. 

\bibliographystyle{plain}
\bibliography{references}

\end{document}